# Strongly Modulated Ambipolar Characteristics of Few-layer Black Phosphorus in Oxygen

Cheng Han[†,1,2,3], Zehua Hu[†,2,3], Jialin Zhang[†,1,2], Fang Hu[1,4], Du Xiang[1,2], Jing Wu[5], Bo Lei[2,3], Li Wang[6], Wen Ping Hu*[7,8], & Wei Chen*[1,2,3,9]

[1]Department of Chemistry, National University of Singapore, Singapore 117543, Singapore

[2]Department of Physics, National University of Singapore, Singapore 117542, Singapore

[3]Centre for Advanced 2D Materials and Graphene Research Centre, National University of Singapore, 6 Science Drive 2, 117546, Singapore

[4]Ningbo Institute of Technology, Zhejiang University, Ningbo 315100, China

[5]Institute of Materials Research and Engineering, A*STAR (Agency for Science, Technology and Research), 3 Research Link, Singapore 117602, Singapore

[6]Institute for Advanced Study and Department of Physics, Nanchang University, 999 Xue Fu Da Dao, Nanchang 330031, China

[7]Beijing National Laboratory for Molecular Sciences, Key Laboratory of Organic Solids, Institute of Chemistry, Chinese Academy of Sciences, Beijing 100190, China

[8]Department of Chemistry, School of Science, Tianjin University & Collaborative Innovation Center of Chemical Science and Engineering (Tianjin), Tianjin 300072, China

[9]National University of Singapore (Suzhou) Research Institute, 377 Lin Quan Street, Suzhou Industrial Park, Jiang Su 215123, China

†These authors contributed equally to this work.

*Authors to whom correspondence should be addressed. Electronic mail: huwp@iccas.ac.cn (W. P. Hu) and phycw@nus.edu.sg (W. Chen)

**Abstract**

Two-dimensional black phosphorus has been configured as field-effect transistors, showing an intrinsic symmetric ambipolar transport characteristic. Here, we demonstrate the strongly modulated ambipolar characteristics of few-layer black phosphorus in oxygen. Pure oxygen exposure can dramatically decrease the electron mobility of black phosphorus without degrading the hole transport. The transport characteristics can be nearly recovered upon annealing in Argon. This reveals that oxygen molecules are physisorbed on black phosphorus. In contrast, oxygen exposure upon light illumination exhibits a significant attenuation for both electron and hole transport, originating from the photoactivated oxidation of black phosphorus, which is corroborated by *in situ* X-ray photoelectron spectroscopy characterization. Our findings clarify the predominant role of oxygen in modulating ambipolar characteristics of black phosphorus, thereby providing deeper insight to the design of black phosphorus based complementary electronics.

**Keywords**: black phosphorus, ambipolar characteristics, field-effect transistors, oxygen, transport modulation

# Introduction

Black phosphorus (BP), as a fast-emerging two-dimensional (2D) material, stands out from other members in 2D family such as graphene[1, 2] and transition metal dichalcogenides (TMDs)[3], and attracts substantial research interests attributed to its remarkably unique fundamental properties and versatile device applications[4-6]. Few-layer BP sheet can be exfoliated from layered BP crystals, where each phosphorus atom is covalently bonded to three neighboring atoms to form a puckered orthorhombic structure[7-9]. BP is featured by a thickness-dependent direct band gap, ranging from ~0.3 eV for bulk to ~2 eV for monolayer[10-12], leading to great potential applications of BP based optoelectronic devices. Moreover, highly anisotropic electronic and optoelectronic characteristics also distinguish BP from most of materials in 2D family[12, 13].

The inherent sizeable band gap enables ultrathin BP to be configured as field-effect transistor (FET) devices, demonstrating a clear ambipolar transport behavior with remarkably high hole mobility up to ~1000 $cm^2V^{-1}s^{-1}$ and on/off ratio of ~$10^5$ at room temperature[14-17]. However, the BP based FETs generally exhibit significant asymmetry between electron and hole transport, where both electron mobility and concentration are orders of magnitudes lower than the hole side, thus seriously limiting its applications in complementary logic electronics. In order to effectively improve the electron transport of BP devices, hence achieving more balanced ambipolar characteristics, several approaches have been utilized such as selection of proper metal contacts[18, 19] and surface transfer doping[20] on BP flakes. Furthermore, R.

A. Doganov et. al report a greatly enhanced electron transport characteristic of pristine few-layer BP channels that are passivated by hexagonal boron nitride in inert atmosphere, compared to the unpassivated and air exposed BP channel[21]. This surface protection process can lead to the appearance of the intrinsic symmetric ambipolar transport behavior of BP, which indicates that air exposure plays a dominant role in suppressing the electron transport. Nevertheless, the key factors to modulate BP ambipolar characteristics in air are still unclear and less understood. A comprehensive spectroscopic investigation (e.g. Raman spectroscopy) has been recently implemented to determine the origin of BP degradation in controlled ambient conditions[22], which reveals the photoactivated oxidation by aqueous oxygen. However, controlled experiments in different ambience have not yet been conducted from FET device perspective. A deeper understanding of how air components (e.g. oxgyen) impact on the BP device performance is very crucial towards BP based complementary electronics.

Here, we report that the ambipolar characteristics of few-layer BP FET devices can be strongly modulated in oxygen ($O_2$). Upon oxygen exposure, the electron transport of BP devices is dramatically suppressed, revealing a decrease of mobility by over three orders of magnitudes; while the hole mobility of BP is nearly retained. Such oxygen induced modulation on electron transport of BP is reversible and the electron transport of BP can be nearly recovered upon annealing under Ar. On the other hand, oxygen exposure under light illumination initiates chemical oxidation of BP, which shows the significant mobility decrease for both electron and hole transport. *In situ* X-ray photoelectron spectroscopy (XPS) investigation further confirms the photoinduced oxidation of BP.

**Results and discussions**

Ultrathin BP flakes were exfoliated from bulk BP crystals and transferred on heavily p-doped silicon substrate coated with 300 nm $SiO_2$, and subsequently configured as two-terminal FET devices for controlled experiments in ambient conditions. Figure 1a displays a typical atomic force microscopy (AFM) image of as-fabricated BP devices. The line profile reveals the BP flake thickness of ~5.4 nm, which corresponds to ~10 atomic layers considering ~0.53 nm interlayer distance in BP crystal. Raman spectrum of the exfoliated few-layer BP (Fig. 1b) demonstrates the characteristic peaks nearly located at 364, 438 and 465 cm$^{-1}$, corresponding to the three main vibration modes in BP, labeled as $A_g^1$, $B_{2g}$ and $A_g^2$, respectively[22, 23].

All the electrical characterizations of as-made BP devices were carried out in high vacuum condition (~10$^{-8}$ mbar). Figure 1c exhibits the typical transfer characteristic ($I_{sd}$-$V_g$) of fabricated BP FETs at $V_{sd}$ = 0.1 V. By applying gate voltage ranging from -80 V to 55 V, the source-drain current increased from OFF to ON state for both negative and positive sweeping, corresponding to the hole and electron transport, respectively. Furthermore, unlike the generally fabricated BP devices in air, the on-current of electron transport in our devices reached the same order of magnitude as that of hole transport, revealing a symmetric ambipolar transport characteristic. The key treatment to obtain such transport behavior is a pre-annealing process of the fabricated devices in inert Argon (Ar) gas filled glove box at 120 °C for more than 30 mins (see Methods). After annealing, BP devices show the more balanced ambipolar behaviors, mainly resulting from the partial desorption of adsorbed air species (e.g. oxygen) on BP surface. Additionally, the inset logarithmic plot shows the current on/off ratio of ~$10^4$, in good agreement with previous reports[13-16]. Extracted from the linear regime of transfer plot, the field-effect mobility of BP flake can be evaluated via the formula below[14, 20]:

$$\mu = \frac{L}{WC_i V_{sd}} \frac{dI_{sd}}{dV_{sd}} \quad (1)$$

where $dI_{sd}/dV_g$ represents the slope of the linear region in transfer characteristic, $C_i$ is the capacitance per unit area between BP and back gate given by $C_i = \varepsilon_0 \varepsilon_r / d$ ($\varepsilon_r$ and $d$ are the dielectric constant and thickness of SiO$_2$, respectively), and $L$, $W$ are the length and width of conduction channel, respectively. For the device in Fig. 1c, the hole and electron mobility were estimated to be on the same order, approximately 83.0 cm$^2$V$^{-1}$s$^{-1}$ and 25.1 cm$^2$V$^{-1}$s$^{-1}$, respectively. The source-drain current versus source-drain voltage characteristics ($I_{sd}$-$V_{sd}$) of the same device, as shown in Fig. 1d, possess excellent linearity for $V_{sd}$ sweeping from 0 V to both 0.1 V and -0.1 V at different $V_g$, revealing good ohmic contacts between BP and metal electrodes.

In order to explore how the adsorbed oxygen influences the transport behavior of BP, fabricated BP FET devices were exposed to purified oxygen at atmospheric pressure, and subsequently evacuated to high vacuum condition for electrical characterizations. Figure 2a demonstrates the typical transfer characteristics evolution of BP devices in logarithmic scale with respect to O$_2$ exposure time. The initial amibpolar transfer curve shows a current minimum nearly located at -27 V. After 640 mins $O_2$ exposure, this minimum gradually shifted along the positive gate voltage to ~-10 V. This suggests a slight p-type doping effect of $O_2$ on BP flake. The carrier concentration of BP induced by a particular $V_g$ in linear region was estimated and plotted with respect to the exposure time (Supporting information Fig. S1). More importantly, the on-current of electron transport in BP was dramatically decreased with increasing exposure time, in particular, by almost three orders of magnitudes after 640 mins exposure; while the hole transport was nearly reserved. This giant attenuation of electron transport with non-degraded hole transport is also clearly illustrated in Fig. 2b, which displays the transfer curve in linear scale after 1280 mins exposure compared to the pristine BP. The calculated electron and hole mobility of BP were plotted as a function of exposure time in Fig. 2c. The electron mobility sharply reduced from 25.1 $cm^2V^{-1}s^{-1}$ to 0.09 $cm^2V^{-1}s^{-1}$ by over three orders of magnitudes after 1280 mins $O_2$ exposure; while the hole mobility almost remained unchanged at ~100 $cm^2V^{-1}s^{-1}$. It is worth noting that the slight increase of hole mobility at the beginning of $O_2$ exposure is mainly ascribed to the insufficient back gate voltage that cannot fully drive the BP device to the linear hole transport regime, thereby limiting the extracted hole mobility at initial exposure stage. In addition, as we further annealed the $O_2$ exposed device in Ar-filled glove box, it was found that the electron transport of BP was remarkably improved (Fig. 2d), and almost returned to the pristine state with the electron mobility of 6.2 $cm^2V^{-1}s^{-1}$. This nearly reversible transfer characteristic of BP suggests that $O_2$ molecules were physically adsorbed on BP surface, such as surface defects[24, 25]. Those physisorbed $O_2$ molecules can serve as electron trapping centers to significantly trap and scatter electron charge carriers during electrical transport. This results in the severely reduced electron mobility as well as the hole doping effect of BP devices. In addition, the pristine BP FET shows an evident hysteresis loop in the transfer characteristic via forward and backward gate sweeping (see supporting information Fig. S2a), which suggests the presence of intrinsic charge trapping sites in BP. Oxygen exposure induced apparently larger hysteresis in BP device, as shown in Fig. S2b-d, thereby leading to the increased density of charge trapping sites in BP due to the oxygen adsorption.

As a comparison, BP based FETs were also exposed to nitrogen ($N_2$). In sharp contrast to the $O_2$ case, $N_2$ exposure did not induce any obvious change in transfer curves of BP FETs with increasing exposure time, as shown in Fig. 3a, thereby giving rise to the almost retained electron and hole mobility upon $N_2$ exposure (Fig. 3b). This further suggests that oxygen plays a predominant role in attenuating the electron transport of BP devices. *In situ* X-ray photoelectron spectroscopy experiments were carried out on $O_2$ exposed bulk BP to further reveal the physical adsorption of oxygen on BP. Figure 3c demonstrates the evolution of P 2p core level XPS spectra of bulk BP as a function of $O_2$ exposure time in dark conditions. Pristine BP exhibits a single 2p peak with spin-orbit split located at the binding energy of ~130 eV, consistent with previous XPS measurements[26-28]. During $O_2$ exposure under atmospheric pressure, we did not observe any obvious change in the evolution of P 2p peak and the appearance of phosphorus oxide related peaks with the binding energy of ~134-135 eV[26, 29, 30]. This clearly excludes the possibility of oxygen induced oxidation of BP.

Inspired by recently proposed photoinduced oxidation of BP in air[22], we further conducted the controlled $O_2$ exposure experiments on BP devices under visible light illumination. Similar to the $O_2$ exposure case, the typical transfer characteristics evolution of illuminated BP devices as a function of exposure time is displayed in Fig. 4a. Here, a 515 nm laser light source with a power intensity of ~1.5 Wcm$^{-2}$ was used to irradiate the BP device in oxygen. The transfer curve of the illuminated BP shows a much faster decrease of electron transport current than the $O_2$ exposed devices, e.g. by four orders upon 640 mins exposure. Surprisingly, the on-current of hole transport was also largely reduced by almost two orders of magnitudes after 1280 mins exposure, in sharp contrast to the intact hole transport in the $O_2$ exposed BP. The transfer curve in linear scale of 1280 mins exposure was plotted with respect to the pristine BP in Fig. 4b, further illustrating the significant suppression for both electron and hole transport of the illuminated BP devices in $O_2$. In Fig. 4c, the hole mobility of BP device progressively degraded from 147.0 cm$^2$V$^{-1}$s$^{-1}$ to 14.5 cm$^2$V$^{-1}$s$^{-1}$ by one order of magnitude upon 1280 mins exposure; while the electron mobility sharply dropped from 21.4 cm$^2$V$^{-1}$s$^{-1}$ down to a negligible value of ~0.007 cm$^2$V$^{-1}$s$^{-1}$ even under 160 mins exposure. Further annealing process partially improved the electron transport of BP device thus reaching the electron mobility of ~0.4 cm$^2$V$^{-1}$s$^{-1}$, as shown in Fig. 4d; while the hole transport remained at the same current level. Such irreversible transport behavior highly differs from the previous $O_2$ exposure case, most likely originating from the photoinduced oxidation of BP. Moreover, the hysteresis of illuminated BP FET was significantly enlarged after $O_2$ exposure (Supporting information Fig. S3), demonstrating the strongly increased charge trapping sites owing to the photoinduced degradation of BP in oxygen.

The light-induced oxidation mechanism of few-layer BP can be expressed as follows:

$$BP + h\nu \rightleftharpoons BP^* \qquad (2)$$

$$BP^* + O_2 \longrightarrow O_2^- + BP + h^+ \longrightarrow PO_x \qquad (3)$$

In equation (2), incident visible light with the photon energy exceeding the BP band gap produces excitons and hence photoinduced electron and hole pairs in BP flake. As shown in equation (3), the adsorbed oxygen molecules can trap those photogenerated electrons to form intermediate superoxide anions, $O_2^-$. The $O_2^-$ and remained photogenerated holes can further induce the oxidation of BP and lead to the formation of phosphorus oxide species, labeled as $PO_x$.

Such photoinduced oxidation mechanism of BP was further confirmed by *in situ* XPS investigations. In contrast to the P 2p spectra evolution of BP upon $O_2$ exposure, light illumination in $O_2$ clearly led to the photoinduced oxidation of BP with a gradual appearance of a phosphorus oxide related peak at the binding energy of ~134.5 eV, as presented in Fig. 4e. The intensity of such $PO_x$ peak progressively increased with the increased $O_2$ exposure and light illumination time. Upon annealing in ultra-high vacuum conditions at 120 °C, phosphorus oxide components cannot be completely removed, revealing the robustness and irreversible nature of the photoinduced oxides of BP. Our results suggest that it is crucial to avoid light irradiation in air environment in order to ensure the high quality and stability of BP based devices.

## Conclusion

We clearly demonstrate the effect of oxygen on modulating ambipolar characteristics of BP FET devices. Oxygen exposure dramatically suppresses the electron mobility in BP FETs by over three orders of magnitudes, while retains a non-degraded hole transport. The degraded electron transport can be nearly recovered after annealing in Ar. In contrast, light illumination in oxygen leads to an obvious photoinduced oxidation in BP and a significant and irreversible attenuation for both electron and hole transport in BP FETs. Our results reveal the dominant role of oxygen in modulating the ambipolar behaviors of BP, thereby facilitating the design of BP based complementary electronic and optoelectronic devices towards practical applications.

## Methods

Few-layer BP flakes were mechanically exfoliated from bulk BP crystals (Smart Elements) onto a degeneratively p-type doped silicon substrate with 300 nm $SiO_2$ using a scotch tape in air. After locating the exfoliated BP flake via a high resolution optical microscope (Nikon Eclipse LV100D), polymethyl methacrylate (PMMA) photoresist was immediately spin coated on the substrate to protect BP from degradation in air. The conventional e-beam lithography technique was subsequently utilized to pattern the source and drain electrodes precisely on the BP flake, followed by the thermal evaporation of 5 nm Ti and 80 nm Au as metal contacts. After liftoff in acetone, the as-fabricate devices were wire-bonded onto a lead chip carrier. The bonded devices were loaded into an Argon gas filled glove box ($O_2$ and $H_2O$ < 0.2 ppm) and subsequently annealed on a hot plate at 120 $^{o}$C for more than 30 mins. The annealed devices were then loaded into a high vacuum system (~$10^{-8}$ mbar) for electrical measurements.

FET characterizations were implemented in a custom-designed high vacuum chamber using an Agilent 2912A source measure unit at room temperature. Highly purified $O_2$ or $N_2$ (> 99.99 %) gas can be introduced into the vacuum chamber through a carefully pumped gas line system. A 515 nm laser light source with the output power of ~11.8 mW (spot diameter ~0.5 mm) was employed to illuminate the sample through a quartz viewport exactly located on top of BP devices. The annealing process of $O_2$ exposed BP devices was conducted in an Ar filled glove box at 120 $^{o}$C for more than 30 mins.

AFM scans of as-made devices were performed in a class 1,000 clean room with controlled humidity of ~50 % using a Bruker Dimension FastScan microscope in tapping mode. Raman spectroscopy measurements were also conducted in clean room via a backscattering configuration using a 532 nm laser as excitation source.

XPS measurements on $O_2$ exposed bulk BP were carried out in a custom-bulit ultrahigh vacuum chamber ($10^{-10}$ mbar) with Mg Kα (1253.6 eV) as excitation sources. Oxygen exposure was undertaken in a load lock chamber with a quartz viewport, and a 532 nm high-power light emitting diode (LED) source of ~1.7 W was used for light illumination (~1 cm × 1cm spot).

## Conflict of Interest

The authors declare no competing financial interest.

## Acknowledgements

Authors acknowledge the technical support from Centre for Advanced 2D Materials and Graphene Research Centre for the device fabrication, and financial support from Singapore MOE Grant R143-000-559-112, National Key Basic Research Program of China (2015CB856505) and NSFC program (21573156).

## Supporting Information Available

**Figure S1:** Carrier concentration of oxygen exposed BP device. **Figure S2:** Hysteresis of the oxygen exposed BP device. **Figure S3:** Hysteresis of the illuminated BP device upon oxygen exposure. This material is available free of charge via the Internet at http://pubs.acs.org.

# Figures

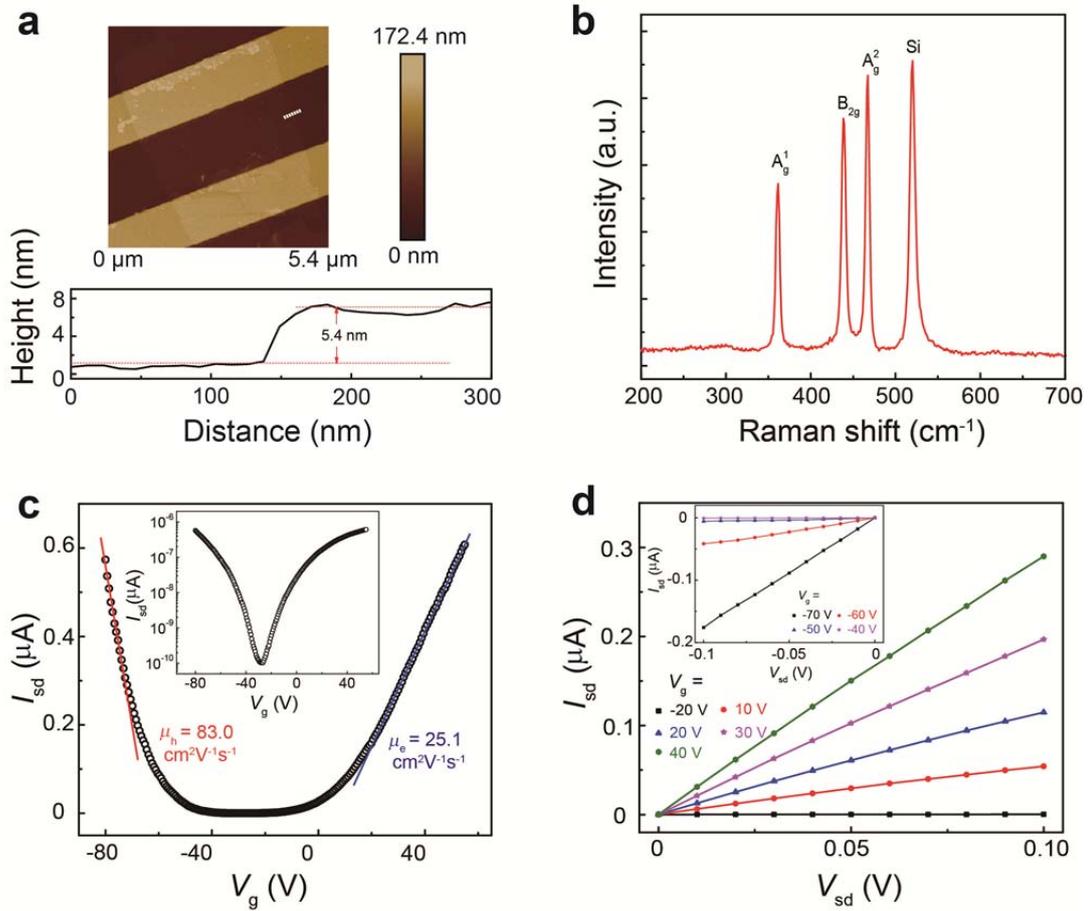

**Figure 1.** **(a)** AFM image of an as-made BP FET device. The line profile indicates a multilayer BP flake of ~5.4 nm (~10 layers). **(b)** Raman spectrum of the BP flake used for device fabrication. **(c)** Transfer characteristics ($I_{sd}$-$V_g$) of a BP FET device at $V_{sd}$ = 0.1 V. Inset: logarithmic plot of the transfer curve. The transfer plot demonstrates a symmetric ambipolar transport characteristic with the hole and electron mobility of 83.0 cm$^2$V$^{-1}$s$^{-1}$ and 25.1 cm$^2$V$^{-1}$s$^{-1}$, respectively. **(d)** $I_{sd}$-$V_{sd}$ characteristics ($V_{sd}$ from 0 V to 0.1 V) of the same device with increasing gate voltages from -20 V to 40 V. Inset: $I_{sd}$-$V_{sd}$ plot with $V_{sd}$ from 0 V to -0.1 V as function of $V_g$ from -40 V to -70 V.

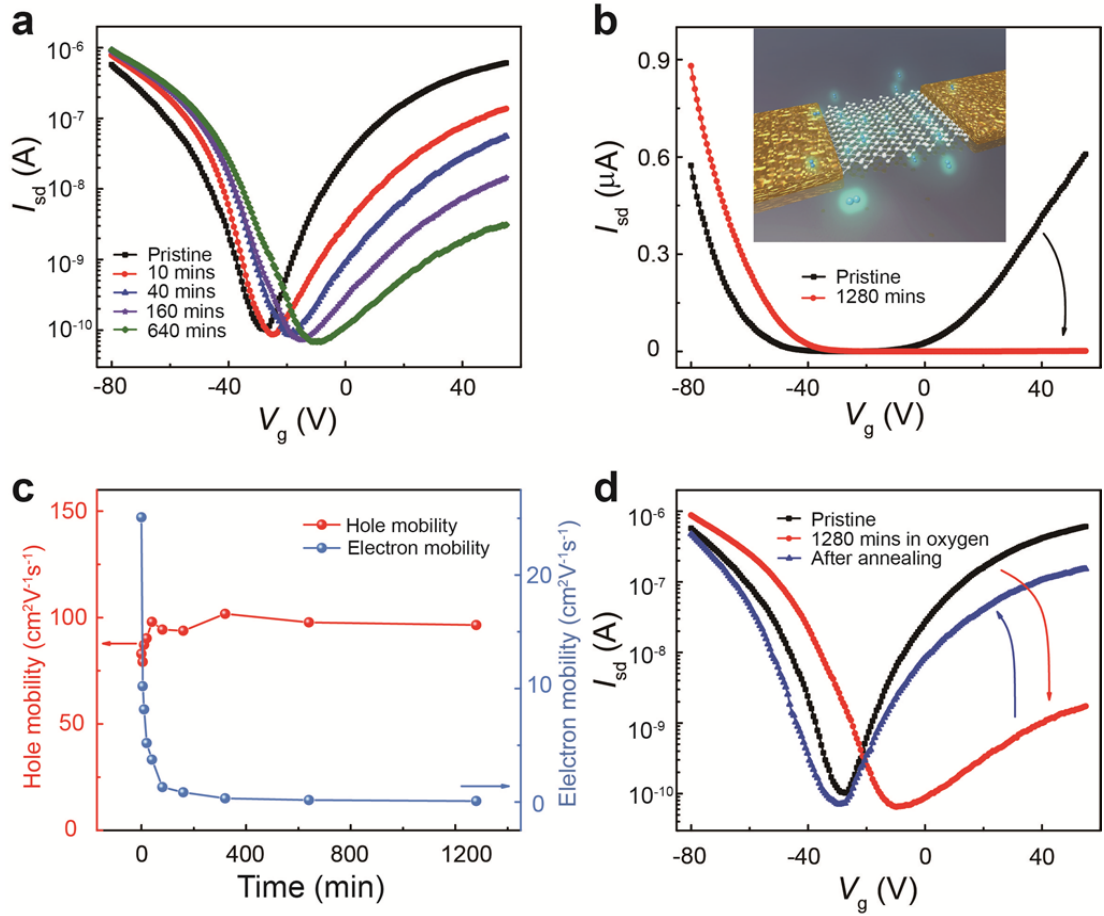

**Figure 2.** (**a**) Transfer characteristics ($V_g$ from -80 V to 55 V) evolution of a BP FET measured at $V_{sd}$ = 0.1 V in logarithmic scale with increasing $O_2$ exposure time from 0 to 640 mins. (**b**) Linear plot of the transfer characteristic upon 1280 mins exposure with respect to the pristine BP. (**c**) The plot of extracted electron and hole mobility as a function of exposure duration. The electron mobility of the BP device is dramatically decreased from 25.1 $cm^2V^{-1}s^{-1}$ to 0.09 $cm^2V^{-1}s^{-1}$; while the hole mobility nearly remained at ~100 $cm^2V^{-1}s^{-1}$. (**d**) Logarithmic plot of the transfer curve after annealing compared to pristine and 1280 mins exposed curves.

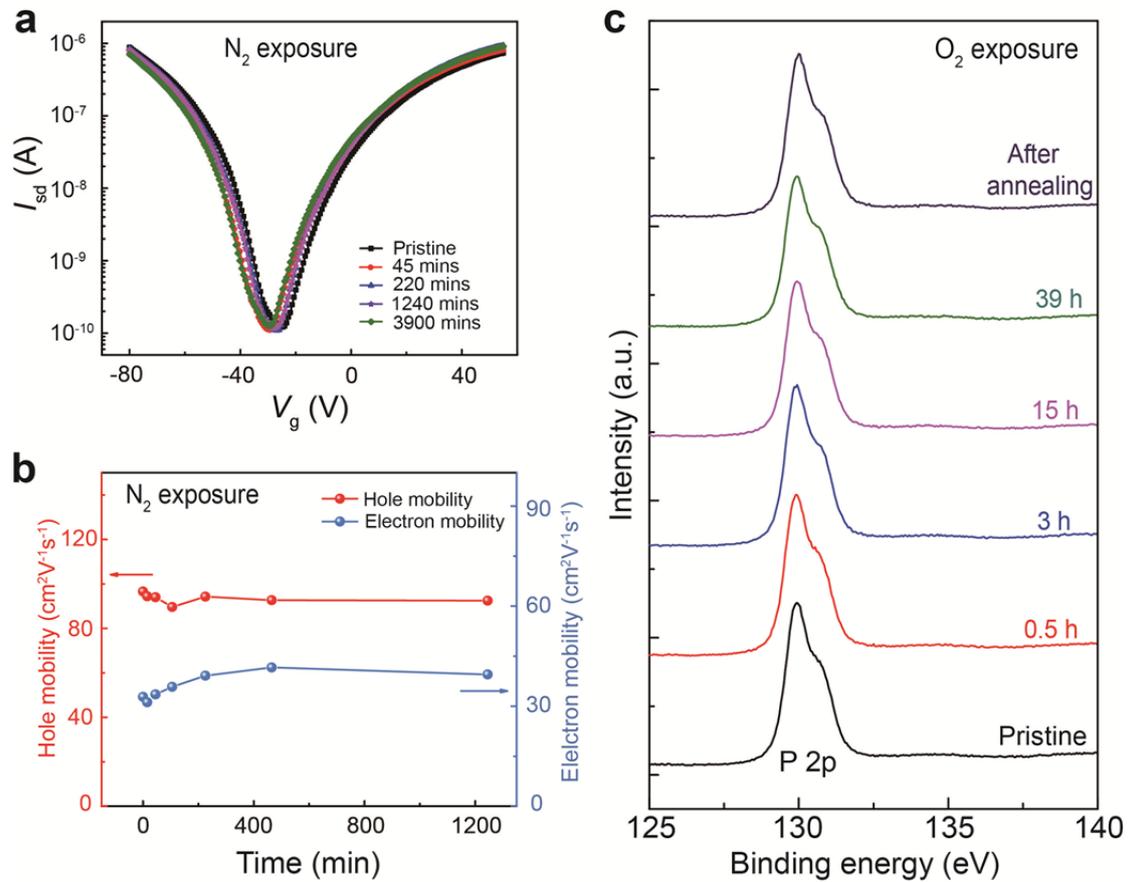

**Figure 3.** (**a**) Logarithmic transfer characteristics evolution (at $V_{sd} = 0.1$ V) of a BP device with respect to $N_2$ exposure time. (**b**) Calculated electron and hole mobility versus exposure time. (**c**) P 2p core level XPS spectra evolution of bulk BP as a function of $O_2$ exposure time in dark conditions.

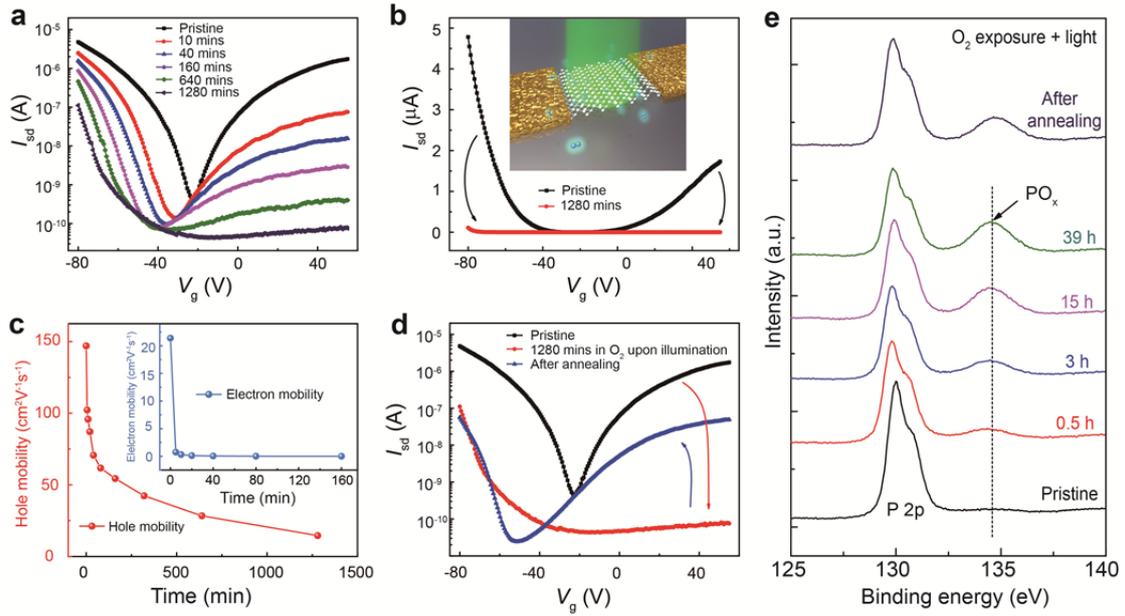

**Figure 4.** (a) Transfer characteristics evolution (at $V_{sd}$ = 0.1 V) of a BP FET in logarithmic scale as a function of $O_2$ exposure upon the illumination of a 515 nm laser (~1.5 Wcm$^{-2}$). (b) Linear transfer plot of 1280 mins exposure with respect to the pristine BP. (c) Extracted electron and hole mobility versus exposure time. (d) The plot of the transfer curve after annealing in logarithmic scale with respect to pristine and 1280 mins exposed curves. (e) The evolution of P 2p core level XPS spectra of illuminated bulk BP upon $O_2$ exposure.

**TOC**

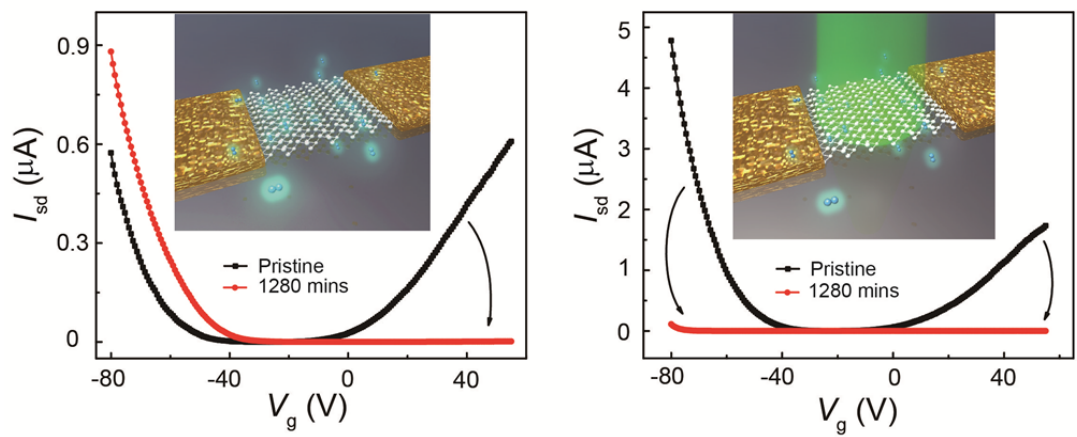

For Table of Content Only

**Supporting Information**

# Strongly Modulated Ambipolar Characteristics of Few-layer Black Phosphorus in Oxygen

Cheng Han[†,1,2,3], Zehua Hu[†,2,3], Jialin Zhang[†,1,2], Fang Hu[1,4], Du Xiang[1,2], Jing Wu[5], Bo Lei[2,3], Li Wang[6], Wen Ping Hu*[7,8], & Wei Chen*[1,2,3,9]

[1]Department of Chemistry, National University of Singapore, Singapore 117543, Singapore

[2]Department of Physics, National University of Singapore, 117542, Singapore

[3]Centre for Advanced 2D Materials and Graphene Research Centre, National University of Singapore, 6 Science Drive 2, 117546, Singapore

[4]Ningbo Institute of Technology, Zhejiang University, Ningbo 315100, China

[5]Institute of Materials Research and Engineering, A*STAR (Agency for Science, Technology and Research), 3 Research Link, Singapore 117602, Singapore

[6]Institute for Advanced Study and Department of Physics, Nanchang University, 999 Xue Fu Da Dao, Nanchang 330031, China

[7]Beijing National Laboratory for Molecular Sciences, Key Laboratory of Organic Solids, Institute of Chemistry, Chinese Academy of Sciences, Beijing 100190, China

[8]Department of Chemistry, School of Science, Tianjin University & Collaborative Innovation Center of Chemical Science and Engineering (Tianjin), Tianjin 300072, China

[9]National University of Singapore (Suzhou) Research Institute, 377 Lin Quan Street, Suzhou

Industrial Park, Jiang Su 215123, China

[†]These authors contributed equally to this work.

[*]Authors to whom correspondence should be addressed. Electronic mail: huwp@iccas.ac.cn (W. P. Hu) and phycw@nus.edu.sg (W. Chen)

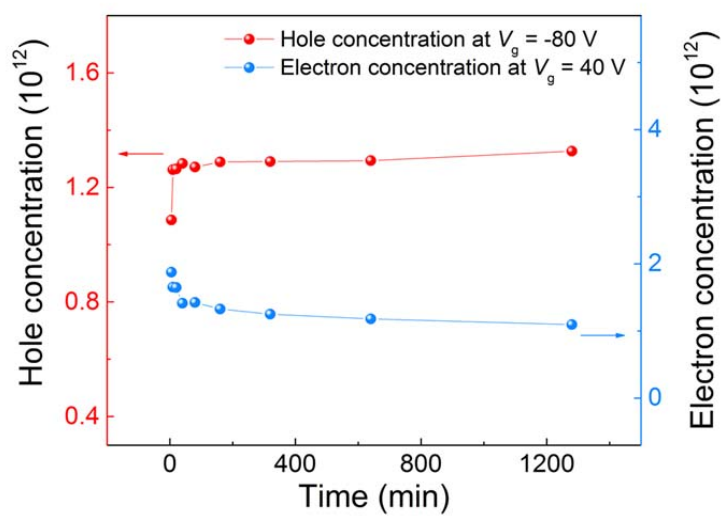

**Figure S1.** Estimated hole concentration at -80 V $V_g$ and electron concentration at 40 V $V_g$ as a function of exposure time. The hole concentration is increased from $9.3 \times 10^{11}$ cm$^{-2}$ to $1.3 \times 10^{12}$ cm$^{-2}$; while the electron concentration is decreased from $2.3 \times 10^{12}$ cm$^{-2}$ to $1.1 \times 10^{11}$ cm$^{-2}$. This reveals the slight p-type doping effect in $O_2$ exposed BP.

On the basis of the transfer curve, the carrier concentration induced by a specific gate voltage $V_g$ in the linear region can be assessed by the following equation:

$$n = -\frac{C_i(V_g - V_{th})}{e}$$

where $C_i$ is the capacitance per unit area between BP and back gate given in main text, and $V_{th}$ represent the threshold voltage that can be extracted from the linear extrapolation of current onset in the linear region of hole or electron side. For example, $V_{th}$ of the ambipolar device in Fig. 1 was determined to be ~-67 V for holes and ~8.5 V for electrons, thereby giving the hole concentration of $9.3 \times 10^{11}$ cm$^{-2}$ at $V_g$ = -80 V and the electron concentration of $2.3 \times 10^{12}$ cm$^{-2}$ at $V_g$ = 40 V.

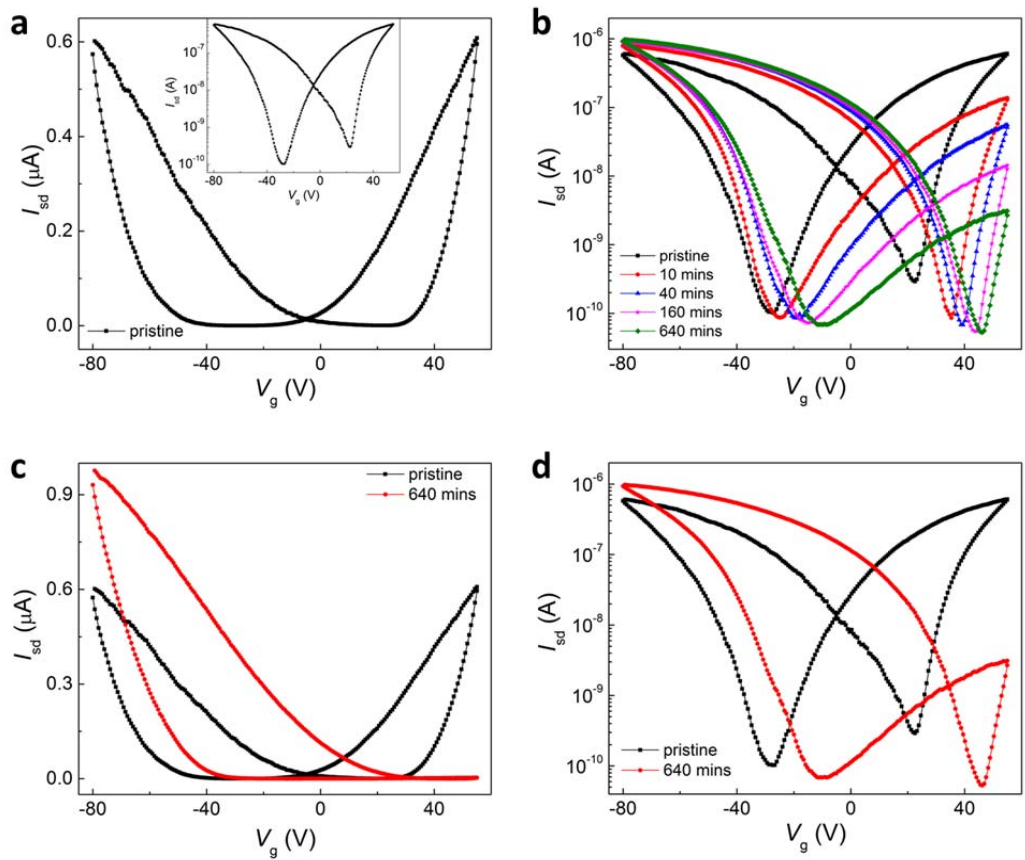

**Figure S2. (a)** Transfer characteristics of pristine BP via forward and backward gate sweeping in linear and logarithmic scale (inset). **(b)** The evolution of bi-directionally gate swept transfer curves with increasing exposure time. The plot of transfer curve upon 640 mins exposure with respect to pristine BP in **(c)** linear and **(d)** logarithmic scale. The hysteresis loop of BP device is clearly enlarged after $O_2$ exposure, revealing the increase of charge trapping sites in BP owing to $O_2$ adsorption.

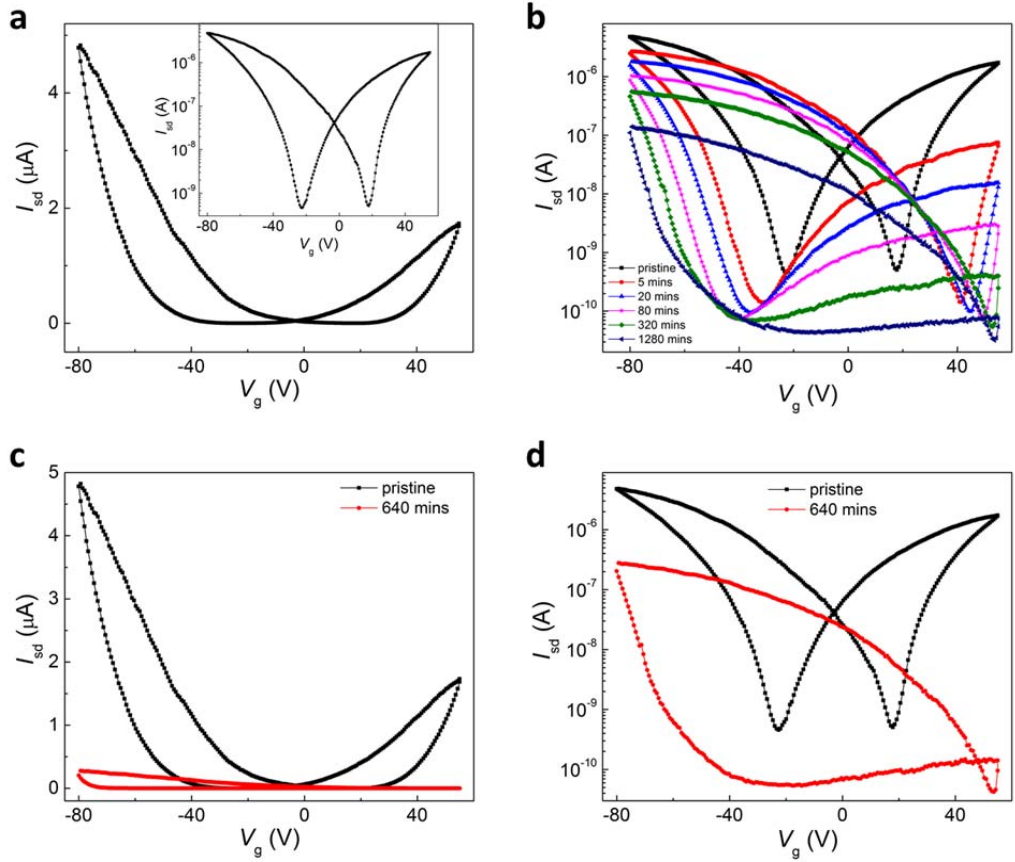

**Figure S3. (a)** Transfer characteristics of pristine BP via bi-directional gate sweeping in linear and logarithmic scale (inset). **(b)** The hysteresis loop evolution as a funcation of exposure time under illumination. The plot of transfer curve upon 640 mins exposure with respect to pristine BP in **(c)** linear and **(d)** logarithmic scale. $O_2$ exposed BP device under irradiation exhibits a significantly increased hysteresis.